\documentclass[preprint,showpacs,preprintnumbers,amsmath,amssymb,amsfonts]{revtex4}

\usepackage{graphicx}% Include figure files
\usepackage{dcolumn}% Align table columns on decimal point
\usepackage{bm}% bold math

\newcommand{\Z}{\mathbb{Z}}

\newcommand{\p}{_{1:t-1}}
\newcommand{\pp}{_{1:t}}
\newcommand{\pn}{_{1:t}}

\begin{document}

%\preprint{APS/123-QED}

\title{On directed information theory  and Granger causality graphs}
\author{Pierre-Olivier Amblard}
 \email{bidou.amblard@gipsa-lab.inpg.fr}
  %\homepage{http://www.lis.inpg.fr/pages_perso/bidou/BidouPerso.html}
 %\altaffiliation[Also at ]{Laboratoire des Images et des Signaux, CNRS, UMR 5083 }%Lines break automatically or can be forced with \\
\author{Olivier J.~J. Michel }%
 \email{olivier.michel@gipsa-lab.inpg.fr}

\affiliation{%
GIPSAlab/DIS, Grenoble, CNRS, UMR 5083\\
BP46 38402 Saint-Martin d'H\`eres cedex , France
}%

\date{\today}% It is always \today, today,

\begin{abstract}
Directed information theory deals with communication channels with feedback. When applied to networks, a natural extension based on causal conditioning is needed. We show here that measures built from directed information theory in networks can be used to assess Granger causality graphs of stochastic processes. We show that directed information theory includes  measures such as the transfer entropy, and that it is the adequate information theoretic framework needed for neuroscience applications, such as connectivity inference problems.

\end{abstract}
\sloppy

\maketitle
\section{Introduction}
\label{intro:sec}
 Modeling and estimating connectivity is a key question often raised in neuroscience. Understanding connectivity  is fundamental in order to decipher how neural networks process information.  Deriving a definition for  connectivity  turns out to be a problem.
 In \citep{Spor07}, three types of connectivities are described: structural or anatomical connectivity describes the physical links between parts of the brain; functional connectivity describes links between parts of the brain that jointly react in some circumstances (the joint reaction is reflected by measures such as correlation or mutual information); effective connectivity is an attempt to add to functional  connectivity the notion of direction in the information flow.
Once a point of view is adopted, the inference problem {\it i.e.} estimating the connectivity from data, gives rise to numerous difficulties.
For instance, in measuring
effective connectivity, the different scales of observation of the brain (associated with different means of observation) lead to time series that may have very different natures and properties, and thus may lead to rather different conclusions.
 When studying, for example,
 networks of neurons cultured {\it in vitro} and recorded by Micro-Electrode Arrays, the
  recorded signals
 will usually be described as a mixture
 of point processes and continuously valued processes. Depending on the nature of the experiment, the correlation structure of the signals may depict short or long memory, leading to different
processing schemes.
 Furthermore, approaches
 will be in general
 highly nonlinear. Going to a much broader scale, fMRI measurements are well modeled by Gaussian processes but with long range memory.
These facts lead to the conclusion
 that there is no universal method for inferring a graph from multiple measurements that will reflect the connectivity  of the brain. However, general principles may be designed and adapted to each situation. It is the goal of this short paper to offer such a general framework---one that relies on
 information theory and causality principles.

 Dependence analysis will provide the main tools for inferring connectivity.
 Such tools
 range from correlation and partial correlation to mutual information and causality measures.
 Many of the most popular tools are non directional, e.g. correlation or partial correlation, and mutual information measures.
These measures have been extensively used in neuroscience ({\it e.g.} \citep{JirsM07,AchaSWSB06,KrasSG04}, to cite but a few).

Alternately, some authors have defined directional measures. Some of these generalize partial correlation to partial directed coherence in order to have efficient second-order statistical methods \citep{KamiDTB01,Eich06}. Other methods
 and measures
 have been developed
 using
 information theoretic tools \citep{SaitH81,Schr00,PaluKHS01,HlavSPVB07}.
 Among these measures, the most popular one, the transfer entropy, is often cited in neuroscience.
  It has been applied, for example, in \citep{LungS06} to measure information flow in sensorimotor
networks.
 Transfer entropy relies, by construction,  on bivariate analysis.
One attempt to generalize it to multivariate analysis has been
 suggested
in \citep{FrenP07}.
 Although not designed for solving neurosciences problem, this method uses a
very interesting and pragmatic approach. We will discuss this in the last section.

A different class of approaches relies on work by Wiener and Granger on causality. Granger causality
considers that a signal $x_t$ causes a signal $y_t$ if the prediction of $y_t$ is increased when taking into account the past of $x_t$. This approach is appealing but gives rise to many questions, philosophical as well as technical \citep{Gran80,Gran88,Gewe82,RissW87,Pear00}. Several levels of definition
for
Granger causality exist. If
the definition based on linear prediction is adopted, operational approaches exist to assess causality between signals.  These approaches and some `linear-in-the-parameters' nonlinear extensions have been applied in neuroscience ({\it e.g.} \citep{Eich05,Seth05,SethE07}).
 Interestingly, applying Granger causality definitions within a linear modeling framework turns out to introduce measures mostly used in correlation based approaches (directed partial coherence). This opens a way to unify the different point of views.

{\em The  goal of the paper is to
 propose
a possible unification  between  Granger causality and information theory.
 This is made possible by recoursing to the
framework of directed information theory }

'Directed information theory'  has its roots in Marko's work; Marko was a German ethologist who studied communication between monkeys in the 1970's \citep{Mark73}.
Marko remarked that standard information theory was not adequate in the context he studied, since feedback was not taken into account by symmetrical quantities such as the mutual information. He thus introduced directed information measures  elaborated from
Markov modeling of
 communication signals. His findings were later (re)formalized by Massey in 1990, developed by Kramers, Tatikonda and some others in the late 1990's, and  more recently
 \citep{Mass90,Kram98,Tati00,VenkP07,TatiM09}.
 All these results and developments may be referred to as
 directed information theory, and culminates in the study of communication theory  through channels with feedback.
Here, we do not consider the problem of communication in its full generality, but
 rather we  consider directed information theory to assess directional dependencies between multiple time series.

 The paper is organized as follows:  Granger causality graphs, as defined by the work in \citep{Eich99,DahlE03}, are introduced in the next section.
Then, we
 present the essentials of directed information theory,
 with emphasis on
  the notion of causal conditioning.
 Causal conditioning
   is fundamental to assess directional dependence between multiple time series.
    While extending these
   tools for stochastic processes,
   we will highlight the relationships
   between transfer entropy and directed information theory \citep{AmblM09:pre,BarnBS09}.
    Section  \ref{info-gran:sec} is dedicated to establishing the link between Granger causality graphs and directed information theory. This is  one of the main points made in this paper.
    Although the
   paper remains deliberately at the conceptual level,
 some practical aspects
   such as estimation issues or testing
 are discussed
     in the last section.

\section{Granger Causality graphs}
\label{grangergraph:sec}

Graphical modeling is a powerful statistical method to model the dependence structure of multivariate random variables \citep{Whit89,Laur96}. Graphical models have been extended to random processes in the nineties \citep{Bril96,DahlES97,Dahl00} and the learning of graphical models have subsequently been studied, {\it e.g.}  \citep{Eich99,Dahl00,BachJ04}.
It is worth noting
 that one of the first  applications  was dedicated to neuroscience \citep{DahlES97}. In \citep{Eich99}, the concept of (linear) causality graph is introduced. Such a graph is a mixed graph in which nodes may be connected by directed edges  as well as undirected edges.
 Each
 connection is defined using the concept of
Granger causality, restricted to linear  models.
 Later, \citep{DahlE03} generalized the definition of connection using the unrestricted Granger causality definition,
 {\it i.e. } based on probability measures.

\subsection{Granger causality}

In this section we briefly review the basics concerning Granger causality between two time series.
Granger causality is based upon prediction theory.
Let $x_t$ and $y_t$ be two stochastic processes indexed by $\Z$, the set of relative integers. Let $x_{n:t}$ be the  vector composed of all the samples of $x$ from time $n$ up to time $t$, or $x_{n:t}= (x_n, x_{n+1},\ldots, x_{t-1},x_t)$. $n$ may be equal to $1$ in which case $x_{1:t}$ represents the whole past and the present of process $x$ at time $t$. We set to $t=1$ the origin of time for the sake of mathematical convenience. Once all the measures are defined, we implicitly let the time origin going to $-\infty$.

 Let capital letters denote multivariate processes, $X_t=(x_{1,t},\ldots,x_{N,t})$. As above,
 $X_{n:t}$ will denote the collection of all the samples of the multivariate time series from time $n$ up to time $t$.

Basically, a signal $x_t$ will be said to
 'Granger cause'
a signal $y_t$ if the prediction of $y_t$ is improved when considering not only its
own
past but
also the past of $x_t$.
Thus a first definition can be given using (conditional) probability measures $P$ of the processes:  $x_t$ does not cause $y_t$ if and only if $P(y_t | y\p, x\p)=P(y_t | y\p)$. In other words, $x_t$ does not cause $ y_t$ if  $y_t$
is,  conditionally to its own past, independent from the past of $x_t$;
the chain $x\p \rightarrow y\p \rightarrow y_t$ is a Markov chain.

This definition may be
satisfactory
 only
if other observations are not taken into account.
 Actually, it has been quoted by Granger that
adding new observations may change the causality relation between two processes, {\it i.e. }
\begin{eqnarray}
P(y_t | y\p, x\p) &\not=&P(y_t | y\p)\nonumber  \\
&\not \Longrightarrow & \nonumber \\
P(y_t | y\p, x\p, Z\pn)&\not=&P(y_t | y\p,Z\pn).
\end{eqnarray}
The dependence relationship between two times series $x$ and $y$ is not guaranteed to be conserved when extra observations are taken into account.
This means that
Granger causality can only  be considered as a property relative to the available information set.

A very simple example  to illustrate this
can easily be constructed.
Let $x_t= a z_{t-1}+ \varepsilon_t$, $y_t=b  x_{t-1}+ \varphi_t$ and $z_t=c y_{t-1} + \eta_t$ be three processes constructed from three independent processes $\varepsilon,\varphi,\eta$. Then
$P(x_t | x\p,y\p ) \not = P(x_t | x\p )$ whereas $P(x_t | x\p,y\p,z\pn )  = P(x_t | x\p,z\pn )$.
 From this example, we may conclude that a relationship exists between $y$ and $x$ if $z$ is not taken into account.
 If the observation of the third signal $z$ is considered as well,
 no direct link from $y$ to $x$
  is exhibited, as all dependencies between $y$ and $x$ appear to be related to the presence of $z$;
 including $z$ in the analysis, $y$ is found to not Granger cause $x$.

Granger causality is thus
 mainly
due to the influence of the past of a process onto the present of another process.
 Geweke \citep{Gewe82} introduced the definition of instantaneous coupling.
 If the dynamical noises $\varepsilon_t,\varphi_t,\eta_t$ in the preceding example are assumed to be white but no longer independent processes,
there
 is a coupling between $x_t$, $y_t$ and $z_t$ which is instantaneous (Eichler uses the word contemporaneous). Thus two types of influence have to be defined.

Let $x_t$ and $y_t$ be two stochastic processes, and $Z_t$ a third multivariate process which does not contain $x$ nor $y$ as components.
\begin{enumerate}
  \item $x_t$ does not cause $y_t $ relatively to $Z_t$  $\Longleftrightarrow $    $P(y_t | y\p, x\p, Z\pp)  = P(y_t | y\p,Z\pp)$, $\forall t>1$
  \item $x_t$ does not instantaneously cause $y_t $ relatively to $Z_t$   $\Longleftrightarrow $    $P(y_t | y\p, x\pp, Z\pp) = P(y_t | y\p, x\p,Z\pp)$, $ \forall t>1$.
\end{enumerate}
The absence of a causal relation from $x_t$ to $y_t$ corresponds to  the independence between the present of $y$ and the past of $x$, conditionally to the past  of $y$ and the extra information ($Z\pp$).
Further,  the lack of instantaneous causality is symmetrical
 with respect to $x$ and $y$, since it simply states that $x$ and $y$ at time $t$ are independent conditionally on their
joint past and on the past of $Z$.

These definitions
 enable us
to construct a graph from a multivariate time series
  as follows \citep{Eich99,DahlE03}.
  Each time series is associated to a node.
Two types of edges may exist between two nodes.  A directed edge
from node $x$ to node $y$ will mean that $x$ Granger causes $y$ with respect to the remaining time series, and an undirected edge
between $x$ and $y$ will mean that $x$ instantaneously causes $y$ with respect to
the other observed time series, stacked in $Z$.
The undirected nature of the latter edge is
 a consequence of the symmetry of instantaneous causality.
Precisely, let $X_t$ be an $M$-dimensional time series, whose components are denoted as $x_{i,t}, i=1,\ldots, M$. Let $(V,E_d,E_u)$ be the associated
mixed graph, where $V$ is the vertex or node set, $E_d$ is the set of directed edges and $E_u$ is the set of undirected edges. The cardinal of $V$ is $M$.  The
 vertices
in $V$ are labelled by $i=1,\ldots, M$, and vertex $i$ will
 correspond
to process $x_i$ unambiguously. Then, the edge sets are defined {\it via}
\begin{enumerate}
\item $\forall i\in V, j\in V, (i,j) \not \in E_d \Longleftrightarrow$  $x_{i,t}$ does not cause $x_{j,t} $ relatively to $X\backslash\{x_{i},x_{j}\}_t)$
\item $\forall i\in V, j\in V, (i,j) \not \in E_u \Longleftrightarrow$  $x_{i,t}$ does not instantaneously cause $x_{j,t} $ relatively to $X\backslash\{x_{i},x_{j}\}_t$
\end{enumerate}
where
$X\backslash\{x_{i},x_{j}\}_t$ is the $(M-2)$-dimensional process constructed from $X_t$ by deleting  components $i$ and $j$.
$(V,E_d,E_u)$  defines a Granger Causality graph.

\section{Directed information theory}
\label{dirinfo:sec}
This section reviews the
  essential tools from directed information theory,
 but
 not from a communication theory point of view.
Our purpose is instead to recast some results and definitions within the framework of dependence analysis between stochastic processes.
 The link between directed information measures and Granger causality graph will be developed in the next
paragraph.

\subsection{Directional dependence between two stochastic processes}
 For the sake of readability, this paragraph focuses upon studying the relations that may occur between two processes only, namely $x$ and $y$. The role played by the existence of other observed process, outlined previously,  and the importance of accounting for such 'extra information' is deferred to a later discussion.

 From a probabilistic point of view, this dependence structure is encoded in the joint probability measures
$P(x_{n_1}, \ldots x_{n_N} ;  y_{n_1}, \ldots y_{n_N} ) $ for all  $N$ and all times $n_1,\ldots, n_2$ in $\Z$. To introduce the different
 definitions,
we restrict the presentation to the dependence between vectors constructed from the time series, {\it i.e. } $x_{1:t}$. The extension to stochastic processes is discussed in section \ref{stocproc:ssec}. Furthermore, we assume in the sequel that the measures are
absolutely continuous with respect to Lebesgues  measure, and we will work with probability density functions.

If there is no dependence structure, or if the processes are independent, it is well known that the joint probability density functions factorize into
$p(x_{n_1}, \ldots x_{n_N} )\times p(  y_{n_1}, \ldots y_{n_N} ) $. Consider the
Kullback-Leibler
divergence $D_{KL}(f||g)=E_f[\log f(x)/g(x)]$, where
$E_f[ . ]$ is the expectation operator (or ensemble average)
with respect to
the probability  density function $f$. The Kullback-Leibler divergence provides a measure of information when wrongly assuming a random variable as distributed from $g$ when it is in fact distributed from $f$. Choosing for $f$ the joint probability density function between two processes, and for $g$ the product of the marginals then leads to a measure of independence, the well-known mutual information
\begin{eqnarray}
I(x\pn ; y\pn ) = E \left[  \log \frac{p(x\pn ; y\pn ) }{p(x\pn ) p( y\pn )}  \right].
\label{infomut:eq}
\end{eqnarray}
Mutual information is a positive quantity (which is a property inherited from the Kullback-Leibler divergence) and is zero if and only if the two processes are independent \citep{Pins64,CoveT93}.
However it suffers from  being symmetrical with respect to  $x$ and $y$ and consequently it is useless when it comes to measuring directionality in the dependence structure.

This symmetrical behavior appears to be closely related to the symmetry of the factorization  of the joint probability density function $p(x\pn ; y\pn ) = p(x\pn ) p( y\pn )$ under the hypothesis that the processes are independent.
Alternately, the following factorization is introduced:
\begin{eqnarray}
p(x\pn ; y\pn ) &=&  \overleftarrow{p} (x\pn | y\pn ) \overrightarrow{p}(y\pn | x\pn)  \\
\overleftarrow{p} (x\pn|y\pn ) &= &  \prod_{i=1}^t p(x_i | x_{1:i-1},y_{1:i-1} ) \\
\overrightarrow{p}(y\pn | x\pn) &= &\prod_{i=1}^t  p(y_i | x_{1:i},y_{1:i-1}).
\end{eqnarray}
 If we consider
 the link between $x$ and $y$ as a channel with input $x$ and output $y$, the term $\overrightarrow{p}(y\pn | x\pn) $
describes the feedforward link whereas  $\overleftarrow{p} (x\pn|y\pn )$ describes the feedback term.
In the absence of feedback in the channel
the input $x$ at time $t$ does not depend on the past of the output up to time $t-1$, and the feedback factor reduces to
$\overleftarrow{p} (x\pn|y\pn ) = p(x\pn)$.

 Mutual information is a divergence measure between the actual joint probability density function and its factorized equivalent expression when independence holds.
 In order to
assess directionality, Massey suggests to compare the joint probability to the alternative factorization $\overleftarrow{p} (x\pn|y\pn ) p(y\pn)$,
which correspond to a situation of no influence of $x$ onto $y$ but of the existence of feedback from $y$ to $x$. A very simple example is
 given by
$x_t=\alpha x_{t-1} + \beta y_{t-1} +v_t$ and $y_t= \gamma y_{t-1} +w_t$ where $v_t$ and $w_t$ are white noises independent from each other.

The directed information is defined as
\begin{eqnarray}
I(x\pn \rightarrow  y\pn ) = E\left[  \log \frac{p(x\pn ; y\pn ) }{\overleftarrow{p} (x\pn | y\pn ) p(y\pn)}  \right].
\end{eqnarray}
Comparing this definition with equation (\ref{infomut:eq})
 it is observed that the difference lies in the term
$p(x\pn )$
 which is
replaced
 here by
 the term $\overleftarrow{p} (x\pn|y\pn )$.
  This shows that
 the directed information and mutual information will be equal when there is no feedback.
The main properties of the directed information are now summarised. In the sequel, the delay operator $D: x_t \longrightarrow x_{t-1}$ is denoted as $Dx_t$ for a signal and $Dx\pp=(0, x_1,\ldots, x_{t-1}) = (0,x_{1:t-1}) $ for a vector. Different proofs of the results presented hereafter exist,
the simplest of which relies on
 the use of Kullback-Leibler divergence properties.
For detailed proofs, refer to \citep{Mass90,Kram98,Tati00,AmblM09:pre}. The properties are as follows.
\begin{enumerate}
  \item The directed information is positive.
  \item The directed information is smaller than, or equal to the mutual information.
  \item Equality between the  directed information and the mutual information occurs if and only if there is no feedback.
  \item The directed information decomposes as
  \begin{eqnarray}
I(x\pn \rightarrow  y\pn ) + I(Dy\pn \rightarrow  x\pn )  = I(x\pn ; y\pn)
\end{eqnarray}
\end{enumerate}
%%%
The first three points are fundamental from a communication point of view. Point 2 and 3 mean  that mutual information overestimates the quantity of information flowing from one signal to another. This has been used by information theorists to provide closer bounds for the capacity of a channel with feedback. The third point
 ensures that directed information theory leads to the usual theory if there is no feedback. The last point is important as it shows how the information shared by two stochastic processes is decomposed into the sum of information flowing in opposite directions.
 A similar decomposition will be found in the sequel, in the framework of causal conditioning. The purpose of the next section is to provide appropriate definitions for causal conditioning and to open new perspectives for directed information.

\subsection{Causal conditioning, causal conditional directed information}
 An alternative formulation for directed information may be easily obtained:
\begin{eqnarray}
I(x\pn \rightarrow  y\pn ) = \sum_{i=1}^t I\left( x_{1:i} ; y_i \big| y_{1:i-1}  \right),
\end{eqnarray}
where $I(x ;y | z)$ is the conditional mutual information between $x$ and $y$ given $z$.
Directed  information may also be expressed as a function of Shannon entropies as
\begin{eqnarray}
I(x\pn \rightarrow  y\pn ) = H(y\pn) -  \sum_{i=1}^t H\left( y_i \big| x_{1:i} ,y_{1:i-1}  \right).
\end{eqnarray}
This expression should be compared to the expression of mutual information below
\begin{eqnarray}
I(x\pn ; y\pn ) = H(y\pn) -  \sum_{i=1}^t H\left( y_i \big| x_{1:t} ,y_{1:i-1}  \right).
\end{eqnarray}
It appears that
the only difference
 lies in the time horizon over which the conditioning is performed in the conditional entropy. For the mutual information, conditioning is performed for each time over the whole observation of $x$. For the directed information, conditioning  for the term at time $i$ is performed from the time origin up to time $i$. Kramers
 suggested referring to this  conditioning as 'causal conditioning'.
 We keep the same name but propose a slightly different presentation for it.
Causal conditional entropy is defined as
\begin{eqnarray}
H(y\pn  || x\pn ) = -E\left[  \log \overrightarrow{p}(y\pn | x\pn)  \right].
\end{eqnarray}
It quantifies the information that remains when observing $y$ once $x$ has been causally observed.
The directed information is then recovered by subtracting the latter quantity from the entropy of $y$:
\begin{eqnarray}
I(x\pn \rightarrow  y\pn ) = H(y\pn) - H(y\pn  || x\pn ).
\end{eqnarray}
Causal conditioning and usual conditioning can be mixed. Kramers proposes the following rule: when reading from left to right, the first type of conditioning is applied. Thus, according to this rule, we define
\begin{eqnarray}
H(y\pn  \big| x\pn \big|\big|  z\pn) &=& H(y\pn  , x\pn \big|\big|  z\pn ) - H(x\pn \big|\big|  z\pn ) \label{classcauscond:eq}  \\
H(y\pn  \big|\big| x\pn \big|  z\pn) &=& \sum_{i=1}^t H(y_i | y_{1:i-1}, x_{1:i},  z\pn) \label{causclasscond:eq}
\end{eqnarray}
These two definitions highlight a non commutative property between classical and causal conditioning. In eq. (\ref{classcauscond:eq}), the definition is similar to the definition of usual conditional entropy as the difference between the joint entropy of $x$ and $y$ and the entropy of $x$ alone. In eq. (\ref{causclasscond:eq}), the conditioning on $z$ is global (compared to the conditioning on $x$ which is causal). In that sense, in this definition, the conditioning variable $z$ is not necessarily a signal synchronous to signals $x$ and $y$. Instead, eq.  (\ref{classcauscond:eq}) does not make sense if $z_t$ is not synchronous with $x_t$ and $y_t$.

Finally,
a causal conditional directed information can be defined.
Mimicking the definition of  conditional mutual information ( $I(x;y|z)=H(y|z)-H(y|x,z)$ ),  causal conditional directed information is defined as
\begin{eqnarray}
I(x\pn \rightarrow  y\pn \big|\big| z\pn ) &=& H(y\pn\big|\big| z\pn) - H(y\pn  || x\pn , z\pn) \nonumber   \\
&=&  \sum_{i=1}^t I\left( x_{1:i} ; y_i \big| y_{1:i-1} , z_{1:i}  \right).
\end{eqnarray}
This quantity will
 be of crucial importance
 when dealing with multivariate time series. Furthermore, it appears in  the sum of two directed information quantities flowing in opposite directions.
 Actually, it can be shown that
\begin{eqnarray}
I(x\pn \rightarrow y\pn) + I(y\pn \rightarrow x\pn)  &=  & I( x\pn ; y\pn) \nonumber  \\ & &+ \, I(x\pn \rightarrow y\pn || Dx\pn).
\end{eqnarray}
In this expression, the term $I(x\pn \rightarrow y\pn || Dx\pn) $ is
named
 instantaneous exchange information  and can be written as
\begin{eqnarray}
I(x\pn \rightarrow y\pn || Dx\pn) &= & \sum_{i=1}^t I\left( x_{1:i} ; y_i \big| y_{1:i-1} , x_{1:i-1}  \right) \\
&=&\sum_{i=1}^t I\left( x_{i} ; y_i \big| y_{1:i-1} , x_{1:i-1}  \right).
\end{eqnarray}
The last equation is obtained since $x_{1:i} | x_{1:i-1}  = x_{i} | x_{1:i-1} $.
 Furthermore, this equation
 illustrates
 that the instantaneous information exchange is symmetrical in the signals
 $x$ and $y$.

The importance of instantaneous information exchange appears also in the following decomposition of the causal conditional directed information.
Recall the following chain rule for the conditional mutual information \citep{CoveT93}
\begin{eqnarray}
I(x,y ; z |w) = I(x ; z | w) + I( y ; z | w,x).
\end{eqnarray}
Applying it to $I\left( x_{1:i} ; y_i \big| y_{1:i-1} , z_{1:i}  \right)$ leads to
\begin{eqnarray}
I(x\pn \rightarrow y\pn ||  z\pn) &=& \sum_{i=1}^t \left(  I\left( x_{1:i-1} ; y_i \big| y_{1:i-1} , z_{1:i}  \right)  \right. \label{termetransentrop:eq} \\
&& +\, \left. I\left( x_{i} ; y_i \big| x_{1:i-1},y_{1:i-1} , z_{1:i}  \right)  \right) \nonumber \\
&=& I(Dx\pn \rightarrow y\pn ||  z\pn) \nonumber \\
&&+ \,  I(x\pn \rightarrow y\pn ||  Dx\pn, z\pn). \nonumber
\end{eqnarray}
Here, the second term is the  instantaneous information exchange causally conditioned by the third time series $z$. Likewise, the decomposition holds for the directed information
\begin{eqnarray}
I(x\pn \rightarrow y\pn ||  z\pn) &=&  I(Dx\pn \rightarrow y\pn ||  z\pn) \nonumber \\& &+
 I(x\pn \rightarrow y\pn ||  Dx\pn, z\pn).
\label{decomp-id:eq}
\end{eqnarray}

\subsection{Rates for stationary processes}
\label{stocproc:ssec}
%%%
All definitions introduced above make sense for  processes that evolve within a finite dimensional phase space.  Extending these definitions to the study of stochastic processes requires some care. Actually the information related quantities (such as entropy) are extensive. If a stochastic process visits a phase space whose dimension increases with $t$, information quantities often diverge linearly as a function of time. Thus  it makes  sense to introduce information rates, as defined below; these definition extend the classical rates found in the literature:
\begin{eqnarray}
I_\infty(x ; y) &= &\lim_{t\rightarrow +\infty} \frac{1}{t} I(x\pn ; y\pn) \\
I_\infty(x \rightarrow  y) &= &\lim_{t\rightarrow +\infty} \frac{1}{t} I(x\pn  \rightarrow y\pn) \\
I_\infty(x \rightarrow  y  || z ) &= &\lim_{t\rightarrow +\infty} \frac{1}{t} I(x \pn  \rightarrow y\pn || z\pn).
\end{eqnarray}
 All limits are assumed to exist, and the previous quantities are named mutual information rate, directed information rate and causal conditional directed information rate, respectively. A fundamental result allows a simpler expression of the rates when the processes are jointly stationary.
 When dealing with discrete valued processes (and with slightly more involvement, continuous random processes), one can establish that, assuming stationarity,  the
 directed information rates can be written as
\begin{eqnarray}
I_\infty(x \rightarrow  y) &= &\lim_{t\rightarrow +\infty}   I(x\pn  ; y_t | y_{1:t-1}) \\
I_\infty(x \rightarrow  y  || z ) &= &\lim_{t\rightarrow +\infty} I(x\pn  ; y_t | y_{1:t-1}, z\pn).
\end{eqnarray}
A proof of the first equality may be found in  \citep{Kram98}; a proof for the second equality can be derived by following the same lines. Extending these equalities to continuous random processes  relies upon the tools
developed in \citep{Pins64,GrayK80,Gray90}. These equalities
extend
the famous
result for the entropy rate
\begin{eqnarray}
\lim_{t\rightarrow +\infty}  \frac{1}{t} H(x\pn) = \lim_{t\rightarrow +\infty}   H(x_t | x_{1:t-1}).
\end{eqnarray}
Interestingly, applying the preceding results to the decomposition of the directed information in eq. (\ref{decomp-id:eq}) leads to
\begin{eqnarray}
I_\infty(x \rightarrow  y) &= & \lim_{t\rightarrow +\infty}   I(x_{1:t-1} ; y_t | y_{1:t-1}) \nonumber  \\ & &
+ \lim_{t\rightarrow +\infty} I(x_{t} ; y_t |x_{1:t-1}, y_{1:t-1}) \\
&=&  I_\infty(Dx \rightarrow  y)  + I_\infty(x \rightarrow  y || Dx),
\end{eqnarray}
 where
$I_\infty(x \rightarrow  y || Dx)$ is the instantaneous information exchange rate. The other term is the limit of $I(x_{1:t-1} ; y_t | y_{1:t-1})$,
which is a particular instance of Schreiber's transfer entropy \citep{Schr00,KaisS02}. We thus name $I_\infty(Dx_{1:t-1} \rightarrow  y)$
the transfer entropy rate. This result, already mentioned in \citep{AmblM09:pre},
allows to recast all results and approaches found in the literature within a unique and simplified framework.
Further, it highlights the fact that stationarity is implicitly present in Schreiber's intuition, and that  instantaneous information exchange between processes is lacking in his work. The decomposition can be easily done for the conditional rates, and
leads to
\begin{eqnarray}
I_\infty(x \rightarrow  y || z) &= &
  I_\infty(Dx \rightarrow  y||z)  + I_\infty(x \rightarrow  y || Dx,z).
\end{eqnarray}
 This provides an implicit definition of
 conditional transfer entropy rate and
 conditional instantaneous information exchange rate. Furthermore, let us mention that in all the preceding discussion,  the conditioning process $z$ can be  a multivariate process.
We are now ready to link directed information theory and Granger causality graphs.

%%%%
\section{Causal information measures to infer Granger causality graphs}
\label{info-gran:sec}
 When
confronted with a multidimensional time series, a fundamental question is to study  its dependence structure.
 The approach investigated here consists of inferring
a graphical model underlying the process
 that is able to account for causal relationships. A good candidate for such a model is a Granger causality graph \citep{DahlE03}.
 Let $X_t$ be the random multivariate process of interest, and $x_1$, $x_2$ two of its components.
Recall that in a Granger causality graph that models a multivariate process $X_t$,
the absence of a directed edge from  nodes $x_1$ to node $x_2$ is equivalent to the conditional independence expressed by
\begin{eqnarray}
\lefteqn{ P(x_{2,t} | x_{1, 1:t-1} , x_{2, 1:t-1},X\backslash \{x_{1},x_2\}\pp )  = }& &\nonumber \\
& & P(x_{2,t} | x_{2, 1:t-1},X\backslash \{x_{1},x_2\}\pp).
\end{eqnarray}
Similarly,
the absence of an undirected edge  expresses the equality
\begin{eqnarray}
\lefteqn{P(x_{2,t} | x_{1, 1:t} , x_{2, 1:t-1},X\backslash \{x_{1},x_2\}\pp )  =} & &\nonumber \\
&&
  P(x_{2,t} | x_{1, 1:t-1} x_{2, 1:t-1},X\backslash \{x_{1},x_2\}\pp).
\end{eqnarray}
In these expressions $X\backslash \{x_{1},x_2\}$ stands for the multivariate process $X$ without components $x_1$ and $x_2$.

The problem of inferring
a graph from the observed data can then be viewed as a problem of assessing Granger causality between ordered pair of nodes, say $x$ and $y$. This is done relative to the remaining nodes of the graph that form the additional observed  process $X\backslash \{x_{1},x_2\}$.

In view of the previous definitions, we need
 measures
 to assess conditional independence on the past and conditional independence between present samples.
 Such measures were defined in the previous sections, within an information theoretic framework.
We can now state the main results of the paper:

Let $(V,E_d,E_u)$ be the Granger causality graph of a multivariate process $X_t$. Then
\begin{enumerate}
\item $\forall i\in V, j\in V, (i,j) \not \in E_d \Longleftrightarrow$  $I_\infty(Dx_i \rightarrow  x_j || X\backslash \{x_i,x_j\}) =0 $
\item $\forall i\in V, j\in V, (i,j) \not \in E_u \Longleftrightarrow$  $ I_\infty(x_i \rightarrow  x_j || Dx_i, X\backslash \{x_i,x_j\})=0$.
\end{enumerate}
 To state it differently, we have the two following assertions:
 \begin{itemize}
 \item Conditional transfer entropy rate is a well adapted measure in order
   to assess Granger causality between two nodes with respect to the remaining
   available set of observations.
    \item  Conditional instantaneous information exchange rate quantifies the instantaneous causality between two nodes
 relative to the other observed time series (recalling that each node of the graph accounts for a time series).
  \end{itemize}
As a corollary, we can state that there is no edge (directed or undirected) between two nodes $i$ and $j$ if and only if the causal conditional directed information rate  $I_\infty(x \rightarrow  y  || X\backslash \{x_i,x_j\}) ) $ is equal to zero.

These assertions were proven in a previous work for the simpler case of Gaussian processes \citep{AmblM09,AmblM09:pre}. In  \citep{BarnBS09} for the case of bivariate Gaussian processes, the author establishes that transfer entropy can be used to assess Granger causality. However, instantaneous causality is  not mentioned by these authors.
 A sketch of a proof for the general case is given below.

Firstly,  let $x$ and $y$ be two processes such that
  $x$ does not cause $y$
   relative to  a third multivariate process $X$ (which does not contain $x$ nor $y$).
 Testing Granger causality relies upon a  Markov chain dependence model $x_{1:t-1} \rightarrow y_{1:t-1} \rightarrow y_t$ where all dependence is considered conditioned on $X_{1:t}$. According to the assumption `$x$ does not cause $y$', we have $I( x_{1:t-1} ; y_t | y_{1:t-1} , X_{1:t} ) =0 $. Therefore, the sum of such terms in equation
 (\ref{termetransentrop:eq}) equals zero as well. This allows us to assert that for processes that are not  `Granger causally' related, the conditional transfer entropy rate is zero.

Conversely, if the rate is zero, since it  is
defined as
the limit of a sum of positive terms,
each individual terms is
necessarily
equal to zero. Then since conditional independence is equivalent to the nullity of the corresponding conditional mutual information, we
may conclude that the processes are not 'Granger causally' related.

The second assertion is shown in the same way.

\section{Discussion}
\label{discuss:sec}

In this paper, we establish that
Granger causality graphs can be obtained using directed information measures. The
emphasis was put on adapted tools for investigating Granger causal relationships, namely
the conditional transfer entropy rate and the conditional instantaneous information exchange rate.
Interestingly, the sum of these two measures constitutes the causal conditional directed information rate.

We  illustrated that directed information theory may be thought as  a fundamental extension of information theory, especially in the case of neuroscience applications.
 Actually,  feedback is a fundamental ingredient  for modeling and studying  of the brain structures at all scales.
 Directed information, as it is presented here, is shown to be an effective tool to assess  connectivity in the brain. It will have fundamental applications in understanding the processing of information and/or coding information in the brain.

 Although these results are satisfactory from a theoretical point of view, some difficulties remain when it comes to develop practical estimators for the different  information related quantities introduced so far. The remainder is devoted to discussing some practical implementation issues related to the inference of a Granger causality graph.

Firstly,  we have to assume ergodicity and stationarity of the signals if we want to estimate the information rates from a single realization of the multivariate process.
The stationarity assumption further simplifies the analysis, since
this assumption simplifies the definition of
 information rates. In the case of real neural data, the stationarity property is usually satisfied over certain time scales only (it is thus highly context dependent). Regarding ergodicity, this assumption is required, as otherwise time averaging  cannot replace ensemble averages, which may lead  to severe practical difficulties for evaluating statistical quantities.

 Secondly, rates are defined as limits and
  in general cannot be evaluated.
  It is thus usual to introduce a finite length observation window, over which the information measures are evaluated. However, this approach replaces limits by finite size samples and does not not warrant that
  the initial conditions are forgotten; it may introduce some
 systematic bias in the analysis,
 as illustrated  for example in
\cite{AmblM09:pre} for the case of information flows between the components of two dimensional AR(1) processes.
Once
the limitation to finite size samples has been accepted, the estimation of
conditional mutual information
quantities
required
has to be performed. Many estimators can be
applied.
Although we will not describe here  the wealth of  mutual information literature (interested readers may find interesting reviews in
\citep{BeirDGM97,Pani03,HlavSPVB07}, and references therein,
it is worth mentioning
recent promising works on the use of $k$-nearest neighbors to estimate entropies and (conditional) mutual information
\citep{KozaL87,KrasSG04,FrenP07,LeonPS08,WangKV09}.
 One of the most attractive features of these
 techniques lies in the fact that they are almost free of parameters like bin sizes or kernel widths. This allows
 to tackle
 a wide variety of situations,  ranging from
 continuous valued processes
to
  point processes, as illustrated in \citep{Vict02}.
  However, some drawbacks include
   the computational burden  and
  the absence of theoretical results for the rate of convergence.
Nevertheless,
      extensive Monte-Carlo simulations have proved the 
   good behavior of these estimators in moderate dimensions (up to 5 or 6) \citep{KrasSG04,FrenP07,AmblZMC08}. Let us also mention an ingenious trick explained in \cite{FrenP07} which consists for the conditional mutual information $I(x ; y | z )$ in conditioning by the time samples of $z$ that share
    as much
   information as possible with $x$. This allows to effectively reduce the dimension. Another
   rarely considered difficulty
      lies in the different natures and properties encountered in neural data. As
   outlined
     in the introduction,
    neural data may behave as point processes, exhibit some long range dependencies and are often non-stationary.
         These properties
     (and lack of properties)
     make the estimation issue very difficult,
  and
the estimation of information measures, despite a lot of beautiful works, remains a challenging field of research. In this respect prospective works may concern the use of approximate measures based on Gram-Charlier or Edgeworth expansion of the densities \citep{MichF96}.

 The second issue met in practice is the detection issue: assuming that some information rate related measure estimate is available, it must be decided whether an edge exists or not within the graph.
This is a
classical
problem of statistical testing theory
 for which the empirical information rate serves as a test statistics.
Theoretically, if it is zero, no edge is placed between the nodes of interest.
As the measure will practically not be zero
we have to
 choose a
threshold over which the measure is decided to be
significantly
 non zero.
 The most popular approach to solve this problem is due  to Neyman and Pearson, and consists of optimizing the test under the constraint that false positive decision errors (making the wrong decision that an edge exists) remain below some constant chosen value, referred to as the test 'significance level'.

 Of course the level is a probability, and
evaluating its value requires a knowledge of the probability density function of the estimated information rate (serving as the test statistics here)
 under the null hypothesis. Since the
 test
 statistics
 used is a very complicated nonlinear transform of the data, this probability measure is hardly known.
 But  the thresholds to apply can be evaluated by using bootstrapping strategies,
   surrogate data or random permutations  \citep{Good05}.
     This is of course  only
   possible at the expense of an increase in computational load.
    Finally,
      the last problem at hand is that of multiple testing that must be correctly handled. It is known that when multiple testing is performed, as is the case when deciding the presence of edges between multiple pairs of nodes,  controlling the level of the test is not easy \citep{LehmR05}.

%
%% For one-column wide figures use
%\begin{figure}
%% Use the relevant command to insert your figure file.
%% For example, with the graphicx package use
% %% figure caption is below the figure
%\end{figure}
%%
%% For two-column wide figures use
%\begin{figure*}
%% Use the relevant command to insert your figure file.
%% For example, with the graphicx package use
%  \includegraphics[width=0.75\textwidth]{example.eps}
%%\caption{Please write your figure caption here}
%\end{figure*}
%\begin{table}
%\begin{tabular}{lll}
%\hline\noalign{\smallskip}
%\end{tabular}
%\end{table}

%\begin{acknowledgements}
%\end{acknowledgements}


\bibliographystyle{spbasic}      % basic style, author-year citations

 \end{document}